\documentclass[default,sn-mathphys-num,iicol]{sn-jnl}
\usepackage{tabularx}
\usepackage{graphicx}%
\usepackage{multirow}%
\usepackage{amsmath,amssymb,amsfonts}%
\usepackage{amsthm}%
\usepackage{mathrsfs}%
\usepackage[title]{appendix}%
\usepackage{xcolor}%
\usepackage{textcomp}%
\usepackage{manyfoot}%
\usepackage{booktabs}%
\usepackage{algorithm}%
\usepackage{algorithmicx}%
\usepackage{algpseudocode}%
\usepackage{listings}%
\usepackage{comment}
\usepackage{subfig}
\theoremstyle{thmstyleone}%
\theoremstyle{thmstyletwo}%

\theoremstyle{thmstylethree}%

\begin{document}

\title[Po]{High-precision measurement of $^{215}$Po half-life via delayed-coincidence analysis}


\author[1,3]{\fnm{L.} \sur{Ascenzo}}\email{lorenzo.ascenzo@student.univaq.it}

\author[2,3]{\fnm{M. H.} \sur{Baiocchi}}\email{melissahoda.baiocchi@gssi.it}

\author[2,3]{\fnm{G.} \sur{Benato}}\email{giovanni.benato@gssi.it}

\author[2,3]{\fnm{Y.} \sur{Chu}}\email{yingjie.chu@gssi.it}

\author[3]{\fnm{G.} \sur{Di Carlo}}\email{giuseppe.dicarlo@lngs.infn.it}

\author[4,5]{\fnm{A.} \sur{Molinario}}\email{andrea.molinario@inaf.it}

\author[4,5]{\fnm{S.} \sur{Vernetto}}\email{vernetto@to.infn.it}

\affil[1]{\orgname{Universit\`a degli Studi dell'Aquila}, \orgaddress{\street{Via Vetoio 42}, \city{L'Aquila}, \postcode{67100}, \country{Italy}}}

\affil[2]{\orgname{Gran Sasso Science Institute}, \orgaddress{\street{Viale F. Crispi 7}, \city{L'Aquila}, \postcode{67100}, \country{Italy}}}

\affil[3]{\orgname{INFN - Laboratori Nazionali del Gran Sasso}, \orgaddress{\street{Via G. Acitelli 22} \city{Assergi (AQ)}, \postcode{67100}, \country{Italy}}}

\affil[4]{\orgname{INAF - Osservatorio Astrofisico di Torino}, \orgaddress{\street{Via Osservatorio 20}, \city{Pino Torinese (TO)}, \postcode{10025}, \country{Italy}}}

\affil[5]{\orgname{INFN - Sezione di Torino}, \orgaddress{\street{Via P. Giuria 1}, \city{Torino}, \postcode{10125}, \country{Italy}}}


\abstract{
  We performed a high-precision study of the $^{215}$Po $\alpha$-decay using a LaBr$_3$ scintillating detector in a low-background environment.
  The $^{227}$Ac intrinsic contamination  in the LaBr$_{3}$ crystal undergoes a decay chain,
  producing the intermediate pair of $^{219}$Rn$\rightarrow^{215}$Po$\rightarrow^{211}$Pb decays.
  The fast time response and good energy resolution of the detector allow for extracting the short half-life of $^{215}$Po
  from the time correlation of the two subsequent $\alpha$-decays by using the delayed coincidence method.
  Thanks to high statistics and a comprehensive uncertainty assessment, we obtain the most precise half-life value to date of $^{215}$Po,
  corresponding to 1.77804 $\pm$ 0.00091(stat.) $\pm$ 0.00067(syst.)\,ms.%(1.7780$\pm$0.0011)$\,$ms.
}




\maketitle

\section{Introduction}\label{sec:intro}

$^{215}$Po is the isotope with the shortest half-life among the members of the $^{235}$U decay series.
Measuring its half-life is crucial for the determination of the activity of other nuclides in the same chain,
i.e. $^{223}$Ra, $^{227}$Th, and $^{227}$Ac~\cite{kossert2015,kossert2019,Takacs:2023bie}.
This is because the $^{215}$Po might decay during the dead time caused
by the detection of the parent decay of $^{219}$Rn, leading to a loss in the overall counting efficiency.
Additionally, the isotope is of great interest to the high-resolution alpha spectroscopy
for benchmarking precision measurements related to nuclear structures \cite{Tortorelli:2024lei}.
Finally, the experimental $^{215}$Po half-life can be used to compare theoretical calculations
and refine nuclear models \cite{Yahya:2021zld}.
The small variation between the measured half-life and the calculations with theoretical models
can lead to a totally different interaction mechanism.
For these purposes, any discrepancy or uncertainty in the half-life values
would affect the precision of experiments as well as theories and should therefore be investigated thoroughly.

%$^{215}$Po has a much shorter half-life than other decay products in the decay series of $^{235}$U.
%This is often an advantage in terms of identifying the decay process
%through their time signature in a low background measurement \cite{Baccolo:2021odk,Ascenzo:2025ujf}.
%With a $Q$-value of 7525$\,$keV, $^{215}$Po has a half-life in the order of milliseconds.
  
Only a few dedicated measurements of the $^{215}$Po half-life are available~\cite{ward1942new, volkov1961, erlik1971, benzoni:2014aka,Takacs:2023bie},
and they are all in reasonable agreement with each other.
A weighted mean of their values is reported as \mbox{(1.781 $\pm$ 0.003)$\,$ms}~\cite{Takacs:2023bie}.
The limitations and challenges of the previous measurements
involve either complex sample preparations or incomplete analysis procedures.
These issues have caused large systematic uncertainties related to the measurement conditions and data analysis,
and thus limited the half-life precision.
The latest and most precise measurement was performed with $^{227}$Ac-loaded liquid scintillators~\cite{Takacs:2023bie}
and exploited the same delayed-coincidence approach used in the present work. 
%This method allows to cancel out the systematics related to any instrumental or analysis efficiency.
The dominant factor affecting their precision is statistical fluctuations.
In order to improve the accuracy of existing half-life values,
further dedicated measurements with higher statistics
and reliable analysis techniques would be beneficial.

LaBr$_{3}$ crystals feature a $^{227}$Ac contamination,
which is difficult to eliminate due to its similar chemical properties
and atomic structure to lanthanum~\cite{milbrath2005}.
The radioactive $^{227}$Ac has a half-life of 21.77~years.
Its decay produces a cascade of daughter nuclei, including the triplet $\alpha$-decay of
$^{223}$Ra$\rightarrow^{219}$Rn$\rightarrow^{215}$Po$\rightarrow^{211}$Pb.
This impurity in LaBr$_{3}$ crystals causes a significant intrinsic background and can be problematic for low level radiation measurements.
However, the contamination allows us to measure the short-lived $\alpha$-decay of $^{215}$Po
by overcoming the self-absorption limitation often associated with the use of an external target.
The LaBr$_{3}$ scintillator is a well-established technology for $\gamma$-ray detection
due to its high light yield of 65000\,photons$/$MeV.
Moreover, its decay time constant of $\sim$15$\,$ns makes the detector suitable for fast timing measurements.
Finally, the light yield and time response of the crystal remain relatively stable against temperature variations~\cite{alharbi2022rejection},
allowing the long-term detector operation without the need for correcting the data acquisition or event reconstruction procedures.

This paper describes a high-precision half-life measurement using a LaBr$_{3}$ detector.
We employ a robust $\alpha$--$\alpha$ delayed coincidence method, which allows us to cancel out the systematics related to any instrumental or analysis efficiency.
This method has previously been demonstrated in nano- and millisecond nuclear decays,
to tag $\alpha$ signals~\cite{pomme2015uncertainty, Azzolini:2021yft}.
The half-life is determined from the time difference between the subsequent decays of $^{219}$Rn$\rightarrow^{215}$Po$\rightarrow^{211}$Pb.
Section~\ref{sec:detdata} describes the experimental setup and data processing.
Section~\ref{sec:analysis} presents the selection criteria used to identify delayed coincidences
and discusses parameters that can potentially bias the half-life determination,
while Section~\ref{sec:discussion} details the half-life evaluation and the associated systematic uncertainties.

\section{Detector and data}\label{sec:detdata}

The detector consists of a cylindrical LaBr$_{3}$ crystal with 5$\,$cm diameter and 5$\,$cm height, coupled to a H2431 photomultiplier tube (PMT)
supplied with $-$1250$\,$V.
We collected waveforms at 250$\,$MHz using a 14-bit CAEN V1725 digitizer.
The recorded pulses have a 400\,ns length, with a 100\,ns baseline.

The delayed coincidence analysis extracts the half-life from the time difference of two consecutive $\alpha$ events
from the $^{219}$Rn$\rightarrow^{215}$Po$\rightarrow^{211}$Pb doublet,
hence it is necessary that the total $\alpha$ count rate is lower than the decay constant of $^{215}$Po.
In other words, the average time difference of $\alpha$ events should be larger than the half-life of $^{215}$Po,
otherwise random coincidence would dominate.
For this purpose, we performed the measurement at the Gran Sasso underground laboratory in Italy,
with a rock overburden of 1400$\,$m to suppress by a factor of $10^6$ the cosmic-ray flux, which could otherwise deteriorate the high-energy spectrum dominated by $\alpha$ events.
In addition, we inserted the detector into a 5$\,$cm thick lead castle to further reduce the environmental background and lower the overall event rate.
As a result, the count rate in the region of interest is dominated by the $\alpha$-decays from the LaBr$_{3}$ crystal.
Since the half-life of $^{215}$Po is much shorter than the other decay products of $^{227}$Ac, selecting $^{219}$Rn and $^{215}$Po decay events is straightforward.

We processed the recorded events offline with the data analysis software \texttt{Octopus}~\cite{octopus},
which computes all pulse-related quantities, e.g. baseline, amplitude, rise- and decay-time,
and converts the integral of the waveform to energy.
We achieved an energy resolution of 2.57\% at 2615$\,$keV using a $^{232}$Th calibration source.
  
Due to a constant energy threshold set in the trigger, variations in baselines or pulse amplitudes lead to the so-called ``time-walk" effect~\cite{regis2012time}.
We corrected this effect by realigning the recorded pulses on the mid-point of the rising edge, obtaining a more precise timestamp evaluation.
More details about the timestamp precision can be found in Section~\ref{sec:discussion}.

In total, we collected 23.5$\,$days of data with an average $\alpha$ event rate of approximately 23$\,$Hz.
Based on the count rate in the region of interest, around 18$\,$Hz, the probability that random coincidences occur in a time window of 5$\cdot$T$_{1/2}$ for $^{215}$Po
is estimated to be around 0.15, yielding a high signal-to-background ratio in the delayed coincidence analysis.
The second fastest decay in the $^{227}$Ac series is $^{219}$Rn$\rightarrow^{215}$Po.
Its 3.96\,s half-life is much longer than the inverse of the count rate.
Additionally, the signature of the $^{223}$Ra$\rightarrow^{219}$Rn$\rightarrow^{215}$Po doublet in the energy spectrum
significantly differs from that of $^{219}$Rn$\rightarrow^{215}$Po$\rightarrow^{211}$Pb.
Therefore, $^{223}$Ra does not represent a significant background contribution.

\section{Analysis and result}\label{sec:analysis}

\begin{figure}[htbp]
  \centering
  \includegraphics[width=\columnwidth]{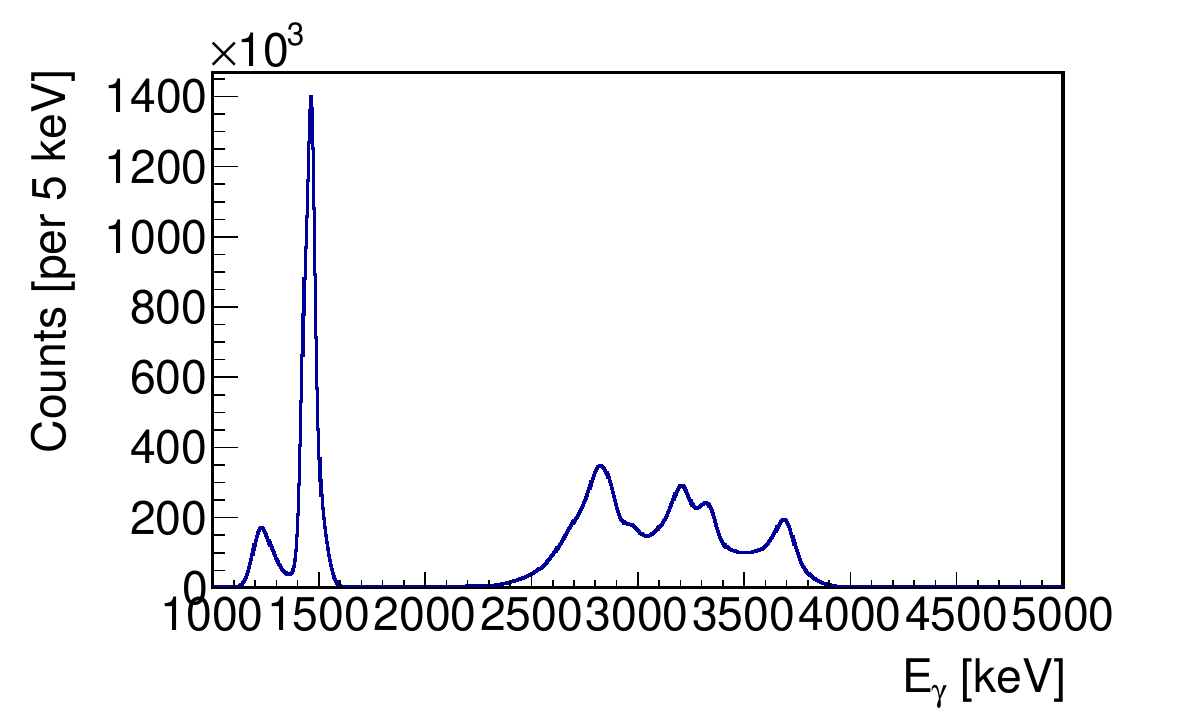}
  \caption{Physics energy spectrum measured for 23.5$\,$days with basic cuts applied.
    A high energy threshold of around 1200$\,$keV was applied during the acquisition.
    The most prominent peak at around 1460$\,$keV is from $^{40}$K contamination.
    The $\alpha$ events lie in separate peaks at higher energies, corresponding to the decay of $^{227}$Ac and its progeny.  }
  \label{fig:Fig.1.0}
\end{figure}

To ensure the data quality of selected events, we applied some basic cuts to the pulse baseline
and the time alignment within the waveform to reject events affected by electronic noise or instrumental instability.
These cuts removed around 0.01\% of the events.
The recorded energy spectrum, calibrated with prominent $\gamma$-rays from the $^{232}$Th source, is shown in Fig.~\ref{fig:Fig.1.0}.
Due to the higher light yield of $\beta/\gamma$ events in the LaBr$_{3}$ crystal,
the $\alpha$ peaks are not reconstructed to their nominal $Q$-values.
However, this has no practical effect on the signal selection for the delayed coincidence analysis.

We attribute the peaks with an E$_{\gamma}$ above 2$\,$MeV to the triplet decay
of $^{223}$Ra$\rightarrow^{219}$Rn$\rightarrow^{215}$Po$\rightarrow^{211}$Pb as identified by the delayed coincidence analysis,
as well as to $\alpha$ decays of $^{227}$Th and $^{221}$Bi in the $^{227}$Ac chain.
These two decay processes have half-life values of several days and are not suitable for the delayed coincidence analysis.
The most energetic peak is composed of $\alpha$'s emitted by $^{215}$Po with a $Q$-value of 7526$\,$keV.
$^{219}$Rn, with a $Q$-value of 6946$\,$keV, exhibits a double-peak feature in the E$_{\gamma}$ range of 3000$\,$keV to 3600$\,$keV,
because the decay can be accompanied by $\gamma$ de-excitations,
which shift the second peak to larger energies due to the higher light yield of $\gamma$-rays.
The lowest $\alpha$ peak at \mbox{E$_{\gamma}$ = 2800$\,$keV} is partly attributed to the $^{223}$Ra decay (\mbox{$Q$ = 5979$\,$keV}).
This is confirmed by the delayed-coincidence analysis applied to the $^{223}$Ra$\rightarrow^{219}$Rn$\rightarrow^{215}$Po doublet,
which provides a half-life value of 3.98\,s, consistent with the expected one. 
To further refine the signal selection, we apply pulse shape discrimination to identify $\alpha$ events.
This approach allows us to minimize the background caused by random coincidences.

For each of the $^{215}$Po candidates with an energy around the $Q$-value of 7526$\,$keV,
we select all previous $\alpha$ decays with energies around the $Q$-value of 6946$\,$keV
and a time difference $\Delta t$ within a predefined interval.
We also perform the event selection forwards in time to have an accurate estimation of the background.
In fact, inverted event pairs can only be caused by random delayed coincidences.
We model the distribution of the reconstructed time difference with an exponential function describing the $^{215}$Po decay
and a flat background $B$ accounting for the random coincidences:
\begin{equation*}
  \label{1.0}
  %  S &= \integral{N_{0}\cdot \lambda \cdot \exp{-\lambda \cdot t}},
  f({t}) = A \exp{\left(-\frac{\mathrm{ln2}}{T_{1/2}} \Delta t\right)} + B\quad,
\end{equation*}
where $T_{1/2}$ is the half-life of $^{215}$Po.

The possible presence of $\gamma$ de-excitations accompanying the $\alpha$ decay of $^{219}$Rn,
results in a partial overlap of the $^{219}$Rn$\rightarrow^{215}$Po and $^{215}$Po$\rightarrow^{211}$Pb energy signatures.
Therefore, it was necessary to optimize the energy selection to maximize the signal-to-background ratio.
For this optimization, we used 24 hours of measurements, equivalent to $\sim5\%$ of the available statistics.
The remaining 95\% of the data was used to determine the half-life of $^{215}$Po.

We varied the energy intervals for selecting the $\alpha$-decay candidates in 20$\,$keV steps.
For the decay of $^{215}$Po, we tuned the intervals in the ranges of
\mbox{E$_{\mathrm{min}}^{\mathrm{1st}}\in[3410,3560]$\,keV,}
and \mbox{E$_{\mathrm{max}}^{\mathrm{1st}}\in[3830,4070]$\,keV}.
For $^{219}$Rn, we adjusted the lower and upper bounds in the \mbox{E$_{\mathrm{min}}^{\mathrm{2nd}}\in[2970,3190]\,$keV},
and E$_{\mathrm{max}}^{\mathrm{2nd}}\in[3290,3560]$\,keV ranges, respectively. 
It is worth noting that the low-energy tail of the $^{215}$Po candidates overlaps
with the high-energy end of $^{219}$Rn candidates.
If the energy selection intervals overlap, some pair of events could be identified as candidates for both positive and negative $\Delta t$ values.
To avoid such double-counting, in the energy range optimization we always impose the condition of
E$_{\mathrm{min}}^{\mathrm{1st}}\geq$E$_{\mathrm{max}}^{\mathrm{2nd}}$.
This is, however, not enough to fully exclude the possibility of a parent-daughter misidentification.
In fact, a true delayed coincidence event could feature a parent falling into the daughter energy range, and vice versa.
This would induce the presence of a small exponential component for negative $\Delta t$ values, which we account for in the fit.

For each combination, we compute the time difference between two consecutive $\alpha$ events
and fill into a histogram with a bin width of 0.05$\,$ms and a fit range of 20$\,$ms,
from which we reconstruct the corresponding signal and background counts.
Considering the count loss caused by the dead time of electronic modules,
the event entries in the first bins near zero, on both the positive and negative sides, are affected.
These bins are therefore excluded from the analysis.
Further details regarding the effect of the dead time are given in Section~\ref{sec:discussion}. 

To optimize the energy window, we evaluate the significance of a Poisson counting experiment, as the distribution of the time difference follows Poisson statistics. The significance $Z$ is expressed as ~\cite{cowandiscovery}:
\begin{equation*}
  \label{1.1}
  Z = \sqrt{\left(2\left((s+b)\mathrm{ln}\left(1+\frac{s}{b}\right)-s\right)\right)}\quad,
\end{equation*}
where $s$ and $b$ represent the number of signal and background events from each selection, respectively.
The optimal energy windows are defined as the ones leading to the maximal significance,
and correspond to E$^{\text{1st}}\in[3034,3455]$\,keV and E$^{\text{2nd}}\in[3455,4071]$\,keV.
Applying the optimal energy window selection, we obtain 307733 signal events, and 152 background events for each 0.05\,ms bin.
We use these numbers, scaling them to match 95$\%$ of the total statistics, as expectation values for the generation of toy data,
which we then use to optimize the $\Delta t$ fit procedure and quantify any possible fit bias connected
to the bin width, the fit range, and fit method.
To produce the toy data, the expected number of counts $\mu$ in each bin [$t_{i,  \mathrm{min}}, t_{i, \mathrm{max}}$] is
\begin{equation*}
  \label{1.2}
  %  S &= \integral{N_{0}\cdot \lambda \cdot \exp{-\lambda \cdot t}},
  \mu_{i} = \int_{t_{i, \mathrm{min}}}^{t_{i, \mathrm{max}}}{f_{s}(t) \, dt} + B_{i}\quad,
\end{equation*}
where $f_{s}(t)=N_{0}\lambda \exp{(-\lambda t)}$ denotes the signal probability density function.
The parameter $\lambda$ represents the decay rate: $\lambda = \frac{\mathrm{ln2}}{T_{1/2}}$.
We generate the toy data by fixing $T_{1/2}$ to the recommended value of 1.781$\,$ms~\cite{livechart}.  

We vary the bin width from 5$\,$\textmu s to 0.5$\,$ms in non-uniform steps.
The optimal value is a compromise between the need to exploit the information embedded in the $\Delta t$ histogram,
which would require the use of small bins, and the computing time required for the fit, which favors larger bins.
For each bin width, we generate $10^{4}$ toy data applying Poisson fluctuations to both the signal and background components.
As for the fit range, we use a maximum $\Delta t$ of 50$\,$ms, covering more than 99.99\% of the decay events from $^{215}$Po.
We find that the statistical uncertainty on the reconstructed half-life is stable up to bin widths of $\sim0.05$\,ms, and increased for larger bins.
As mentioned above, the events with $\Delta t\lesssim500$\,ns are undetected due to dead time, hence the histogram bins near $\Delta t=0$ have to be removed from the analysis.
On the other hand, the first bins for $\Delta t>0$ are also the ones with the highest signal statistics.
As a result, we choose a 0.02\,ms bin width as it falls in the optimal range in terms of statistical uncertainties on the half-life,
and at the same time, it imposes the removal of a small number of signal events at small $\Delta t$ values.

The fit range also affects the fit precision.
Enlarging the fit range above $5\cdot$T$_{1/2}$ would not increase the signal statistics, but allow to put a stronger constraint on the flat background.
For this reason, we vary the fit range from $\pm$10\,ms to $\pm$50\,ms in 2\,ms steps.
As expected, the resulting statistical uncertainty on the reconstructed half-life does not change significantly.
For the final analysis, we choose the $\pm$30\,ms interval as a fit range. 

Another factor to take into account is the fit method.
We use a frequentist fit with a binned Poisson likelihood performed with \texttt{ROOT}~\cite{root} as the reference method,
with uncertainties obtained from a log-likelihood scan.
In addition to the frequentist fit, we also perform a Bayesian analysis using the \texttt{BAT} software~\cite{bat}.
As expected, given the very high statistics, the two approaches provide the same numerical results.

\begin{figure}[htbp]
  \centering
  \includegraphics[width=\columnwidth]{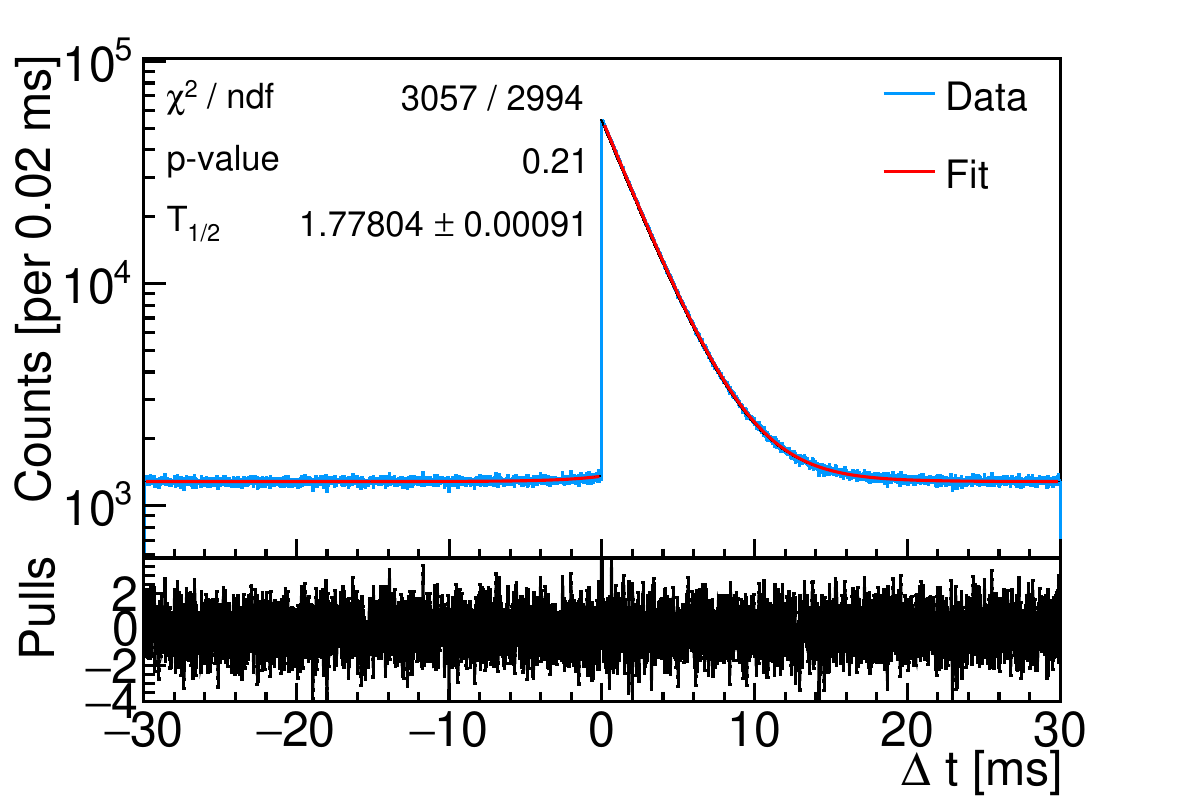}
  \caption{$\Delta t$ distribution  with the exponential fit plus a constant background to determine the half-life $T_{1/2}$.
    %The bottom panel shows pulls defined as the residual in counts divided by the statistical uncertainty in each bin.
  }
  \label{fig:Fig.1.1}
\end{figure}

Once all analysis parameters are fixed, we apply the analysis to the remaining 95\% of the data.
The best fit for the time difference distribution is shown in Fig.~\ref{fig:Fig.1.1}. 

The fit has a reduced $\chi^{2}$ of 1.02 (3057/2994) with a p-value of 0.21.
The best estimate for the half-life is (1.77804 $\pm$ 0.00091)\,ms, where the uncertainty is statistical only.

\section{Systematic Uncertainties}\label{sec:discussion}

When conducting high-precision measurements, it is crucial to carry out a thorough review of any potential systematic errors to ensure the reliability of the result.
This section examines the various instrumental and data analysis factors that could introduce systematic uncertainties
with a magnitude comparable to that of the statistical uncertainty.

Regarding the instrumental effects, one primary source of uncertainty is the time resolution,
which directly influences the half-life reconstruction.
To evaluate the time resolution, we performed dedicated measurements by injecting pulser-generated events, with an amplitude similar to that of $\alpha$ events, directly into the digitizer.
The injected events consist of bursts of two pulses every 5\,ms.
For each burst, a time difference of 1\,ms is set between the two pulses, a value of the same order as the half-life.
These injected events are recorded with the same waveform length as for the physics data.
The reconstructed time difference between two consecutive events at 1\,ms follows a Gaussian distribution
contributing to the $T_{1/2}$ measurement with a systematic uncertainty of 107\,ns.  

We then cross-checked the validity of the result against possible effects induced by the dead time of the digitizer.
In this measurement, we maintained the time interval of each burst at 5\,ms,
while varying the time difference between the two pulses in the burst from 1\,ms down to 400\,ns.
When the time difference is large, all events are expected to be triggered, and the time difference can be well reconstructed.
However, with a small time difference, the second event in a pair may not be recorded due to dead time.
The measurement shows that with a time difference of 435\,ns all events are still triggered,
whereas only 6.3\% of events are triggered with a time difference of 420\,ns.
By considering the digitizer's clock resolution, events with up to 500\,ns time difference might still be affected by this dead-time-induced inefficiency.
However, our choice of neglecting the first bins in the $\Delta t$ distribution, corresponding to $\pm$0.02\,ms, completely bypasses this issue.

Another possible source of systematic uncertainty is the timestamp precision, as mentioned in Section~\ref{sec:detdata},
where a correction procedure is described.
In principle, the inaccuracy may lead to fluctuations in the time difference distribution.
However, the timestamp precision is at the order of 100\,ns,
which is negligible compared to the statistical uncertainty of the result.
We repeated the analysis without applying the timestamp correction,
and found that the reconstructed half-life stays unchanged, as expected.

Furthermore, the $^{215}$Po decay rate should remain stable through the whole measurement period
due to the relatively long half-life of $^{227}$Ac.
However, factors such as temperature and humidity can affect the detector and electronics, adding complexity to the experimental conditions.
To investigate the signal stability over time, we divided the data into 12-hour subsets
and reconstructed both the signal rate and the half-life for each subset.
The distributions of these parameters around their averages are consistent with random fluctuations.
The average half-life is 1.77807\,ms. We consider the difference of 0.00003\,ms between the mean value and the reference fit
as an additional systematic uncertainty.

Any change in detection or reconstruction efficiency is expected not to distort the $\Delta t$ distribution used for the delayed coincidence analysis,
as any inefficiency would reflect into a pair of events not entering the $\Delta t$ plot.
This should be independent of their time difference, except for very close events that could be affected by the dead time.
In any case, we evaluate the uncertainty related to the event selection,
and vary the energy selection criteria in the same ranges as used for the analysis optimization performed on the 5\% dataset.
As a result, we find that the reconstructed half-life values differ from the reference fit by at most 0.00021\,ms. 

Moreover, we account for possible background fluctuations, as the signal is expected to be well-separated from random coincidences.
We evaluate this by either leaving the background as a free parameter or by fixing it with the value extracted from an outer region,
such as in the region of \mbox{[-50, -40]\,ms}. For a conservative estimation, the maximum change relative to the reported half-life is used as the uncertainty,
leading to an uncertainty of 0.00014\,ms. 

Additionally, we consider the uncertainties introduced by the bin width and the fit range.
These two parameters affect the $\chi^{2}$ test. Therefore, we take into account the $\chi^{2}$ variance.
Within a $\chi^{2}$ variation of 0.1, the maximum deviation in the half-life is used to estimate the uncertainty.
We obtain a systematic of 0.00059\,ms for the bin width, and 0.00014\,ms for the fit range.

Finally, we validate the dependence on the fit procedure by performing a least-square fit in \texttt{ROOT} and a Bayesian analysis with \texttt{BAT}.
The fit results of $(1.77803\pm0.00092)$\,ms from \texttt{ROOT} and $(1.77807\pm0.00091)$\,ms from \texttt{BAT}
are in excellent agreement with the reference value, with a maximum difference of 0.00003\,ms.

\begin{table}[htbp]
  \centering
  \caption{Systematic uncertainties resulting from instrumental effects and analysis procedures.
    The total uncertainty is calculated by summing all systematics in quadrature.}
  
  \begin{tabular}{lccccc}
    \toprule

    & Source & Uncertainty [$10^{-4}\,$ms] \\

    \midrule
    \multirow{4}*{Instrumental}   &  Time resolution &  1.1 \\
    &  Dead time &  0 \\
    &  Timestamp &  0 \\
    &  Time stability &  0.3 \\

    \midrule
    \multirow{5}*{Analysis}   &  Event selection &  2.1 \\
    &  Background &  1.4 \\
    &  Bin width &  5.9 \\
    &  Fit range &  1.4 \\
    &  Fit method & 0.3  \\
    \midrule
    Total        &  &   6.7 \\

    \bottomrule
  \end{tabular}
  
  \label{table 1.1}

\end{table}

%In conclusion, any possible bias related to the evaluation of the time difference is negligible
%compared to the statistical uncertainty. Therefore, the measurement precision
%is not limited by systematic uncertainties but by statistical ones.
A summary of all systematic uncertainties is reported in Tab.~\ref{table 1.1}.
Combining all the components in quadrature we obtain a total systematic error of 0.00067\,ms.
For further validation, we performed two completely independent analyses,
obtaining results in full agreement with each other.

\section{Summary and Conclusion}\label{sec:conclusions}

The final result on the half-life of the $^{215}$Po $\alpha$-decay is
\begin{equation*}
  T_{1/2}=1.77804\pm0.00091\,(\text{stat.})\pm0.00067\,(\text{syst.})\,\text{ms}.
\end{equation*}
The systematic error is approximately 75\% of the statistical error, which means that the uncertainty is dominated by the statistical fluctuation, but adding more statistics would not significantly improve the result.

\begin{figure}[htbp]
  \centering
  \includegraphics[width=\columnwidth]{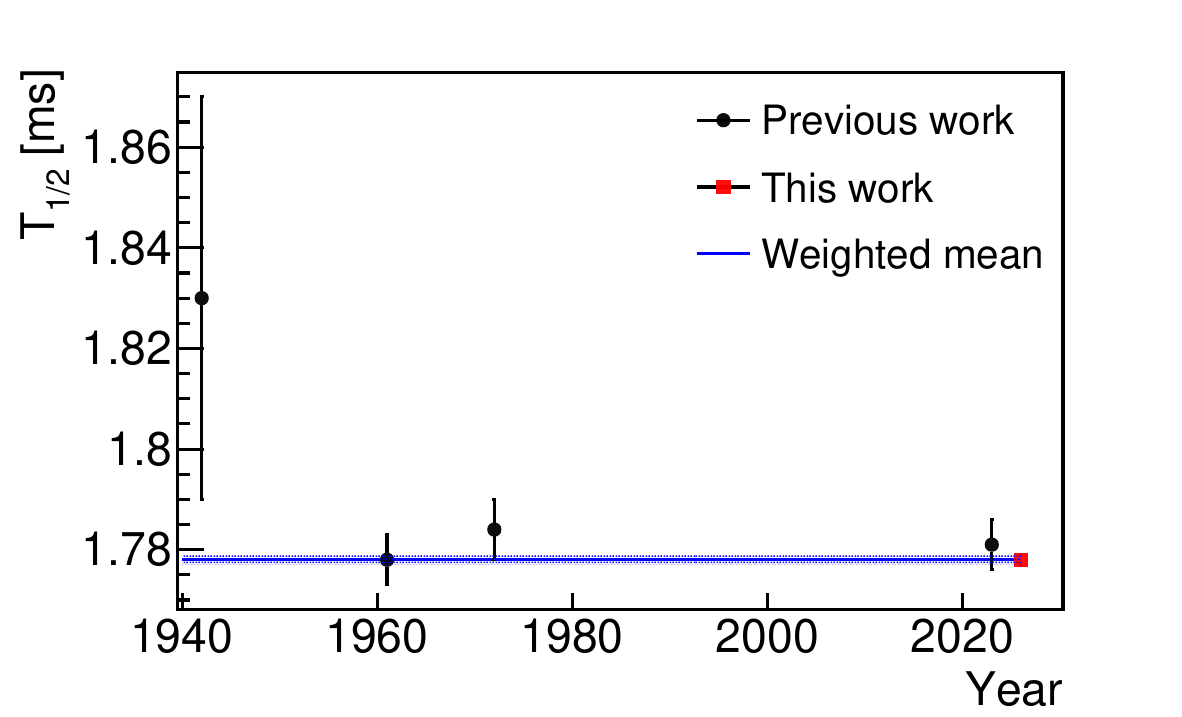}
  \caption{Half-life values taken from Ref.~\cite{Takacs:2023bie} as a function of publication year.
    The black circular points represent the previous published results.
    The red square indicates the value obtained in this work, with the error bar hidden by the marker.
    The weighted mean is calculated by using the method described in Ref.~\cite{rajput1992techniques}.
    The uncertainties associated with the weighted mean are shown in the shaded area. }
  \label{fig:Fig.1.2}
\end{figure}

Figure~\ref{fig:Fig.1.2} shows the previous results and our measured value,
including error bars that display the combined statistical variation and systematic uncertainty.
The weighted mean of 1.778(1)\,ms is derived from all the values illustrated in Fig.~\ref{fig:Fig.1.2}
and is evaluated using the Rajeval technique \cite{rajput1992techniques}.
In comparison, the current half-life value is consistent with earlier measurements and aligns with the weighted mean.
It can be clearly seen that the current measurement has the smallest uncertainty.
However, a slight deviation from the value obtained in 1942 is observed.
One possible source of this discrepancy is attributed to the design of their experimental setup, as already discussed in Ref.~\cite{Takacs:2023bie}.

Employing the delayed coincidence method, we have obtained the most accurate measurement of the half-life of the $^{215}$Po $\alpha$-decay to date.
%corresponding to 1.77804 $\pm$ 0.00091(stat.)$\pm$ 0.00067(syst.)\,ms.
The combined statistical and systematic uncertainty has been improved by a factor of $\sim$4 compared to the latest measurement~\cite{Takacs:2023bie}.
This result is achieved by exploiting the intrinsic $^{227}$Ac contamination of the LaBr$_{3}$ crystal.
Uncertainties from instrumental effects and analysis procedures are studied in details
through dedicated pulser measurements and generated toy data without underestimating factors that may influence the result.
% The combined statistical and systematic uncertainty has been improved by a factor of $\sim$4 compared to the latest result~\cite{Takacs:2023bie}.
This technique has significantly simplified the experimental procedure in comparison with older measurements
and can also be applied to other precision measurements of short-lived $\alpha$-decays.

\backmatter

\bmhead{Acknowledgements}

This work was supported by the European Union, Next Generation EU, Mission 4 Component 1, CUP 2022WWRZZP\_001.

\bibliography{bibliography}% common bib file
%\input{GAGG.bbl}

\end{document}